\documentclass[Afour,times,doublespace]{article}
\usepackage{moreverb,url}
\usepackage{pdfpages}
\usepackage{listings}

\usepackage[switch]{lineno}
\usepackage{lineno}
\usepackage[colorlinks,bookmarksopen,bookmarksnumbered,citecolor=blue,urlcolor=blue]{hyperref}
\newcommand\BibTeX{{\rmfamily B\kern-.05em \textsc{i\kern-.025em b}\kern-.08em
T\kern-.1667em\lower.7ex\hbox{E}\kern-.125emX}}

\usepackage[margin=1in]{geometry}
\usepackage{setspace}

\usepackage{lmodern}        
\usepackage[T1]{fontenc}
\usepackage[utf8]{inputenc}
\usepackage{amsmath,amssymb}
\usepackage[round]{natbib}
\begin{document}

\title{Integrating Knowledge Graphs and Visualization Dashboards for Advance Data Discovery in VESA}

\author{Pawandeep Kaur Betz$^{1}$\thanks{pawandeep.kaur-betz@dlr.de} \\ Tobias Hecking$^{2}$ and Andreas Gerndt$^{1,3}$}






\date{}
\maketitle

\begin{center}
$^{1}$ German Aerospace Center, Department of Visual Computing and Engineering (SC-VCE), Institute of Software Technology, Braunschweig, Germany \\
$^{2}$ German Aerospace Center, Department of Intelligent and Distributed Systems (SC-IVS), Institute of Software Technology, St-Agustin, Germany \\
$^{3}$ University of Bremen, High Performance Visualization Group, Bremen, Germany \\
\end{center}

\section*{Author Information}
\textbf{Dr. Pawandeep Kaur Betz} received her doctorate in data visualization in 2021 from the Friedrich Schiller University of Jena, Germany. Where she was also a research scientist for 6 years in the biodiversity data management project. Since 2022, she is employed as a postdoctoral scientist at the German Aerospace Center (DLR) in the Institute for Software Technology. She primarily works on applied visual analytics for research data management, digital twins and spatial-temporal data. 

\textbf{Dr. Tobias Hecking} completed his PhD in computer science in 2016. Since 2020, he has been employed as head of the "Intelligent Software Systems" working group at the DLR Institute of Software Technology. The aim of his work is to research theories, methods and tools of artificial intelligence for the development of data-driven intelligent software systems. The focus here is on methods of network analysis, natural language processing and machine learning. The main areas of application are knowledge-based software systems, analysis of complex processes and intelligent user interfaces.

\textbf{Prof. Dr. Andreas Gerndt} received his
degree in computer science from Technical University, Darmstadt, Germany
in 1993. In the position of a research
scientist, he also worked at the Fraunhofer Institute for Computer Graphics
(IGD) in Germany. Thereafter, he was a
software engineer for several companies
with focus on Software Engineering and
Computer Graphics. In 1999 he continued his studies in Virtual Reality and Scientific Visualization
at RWTH Aachen University, Germany, where he received
his doctoral degree in computer science. After two years of
interdisciplinary research activities as a post-doctoral fellow
at the University of Louisiana, Lafayette, USA, he returned
to Germany in 2008 to head a department at the German
Aerospace Center (DLR). Since 2019, he is also Professor
in High-Performance Visualization at University of Bremen,
Germany.

\section*{Author Contribution}
\textbf{Pawandeep Kaur Betz}:  Conceptualization, Funding acquisition,
Investigation, Project administration, User Survey and analysis, drafting the paper, writing and revising, accountability of this work \\
\textbf{Tobias Hecking}: Funding acquisition, Investigation, Writing \\
\textbf{Andreas Gerndt}: Final approval of the paper version

\section*{Abstract} 
The increasing complexity and scale of scientific datasets demand advanced tools for efficient discovery and exploration. Traditional search systems often fall short in addressing the multidimensional nature of data and their intricate relationships, limiting their utility for researchers. This paper introduces the Knowledge Graph Based Visualization Search Application (VESA), that reshapes the process of data discovery by leveraging knowledge graph technology to establish meaningful connections and employing a visualization dashboard to enable multidimensional exploration. A software prototype is developed, showcasing our use case of connecting two Earth System Science repositories via a knowledge graph backend and visualization dashboard at the frontend. The framework’s effectiveness was assessed against guidelines derived from a comprehensive literature review and further validated through an online user study. The evaluation revealed positive reception, highlighting VESA’s low learning curve, ease of use, and potential to enhance data discovery workflows.

\section{Introduction}
\label{section:introduction}
Modern data accumulation techniques have led to the rapid growth of datasets, and consequently, data repositories have expanded significantly. 
The FAIR data principles (\cite{wilkinson2016fair}) emphasize the importance of making data ’Findable’ and ’Accessible’. However, this becomes challenging with large repositories containing datasets from various interconnected domains and in different formats. This further inhibits data accessibility for the scientists and ultimately impact their research (\cite{tenopir2011data, holzner2009first}).

Traditional search environments rely heavily on keyword or full text search, as revealed by a survey on 98 data repositories by Khalsa et al. (\cite{khalsa2018survey}). Moreover, usually same criteria is used for data search as for document search. Numerous studies (\cite{kern2015there, kramer2021data}) have repeatedly highlighted the fundamental differences between these
two types of searches, underscoring the need for novel approaches and tailored solutions for data search (\cite{borst2020patterns}).

Furthermore, typical data search environments are designed to support two types of search tasks (\cite{borst2020patterns}) i,e., lookup and exploratory tasks. However, they often overlook the importance of providing an overview of the collection. An overview task is extremely important as it provides a quick feedback on the user queries while presenting a high-level view of the filtered search space. Additionally, for data managers and administrators, it provides a direct snapshot of their collection and consequently also support in the overall quality assurance of the repositories. Another crucial search task that is often neglected is the representation of both implicit and explicit relationships among metadata entities, which can further enhance understanding and discovery.

\begin{figure} [t]
   \includegraphics[width=\linewidth]{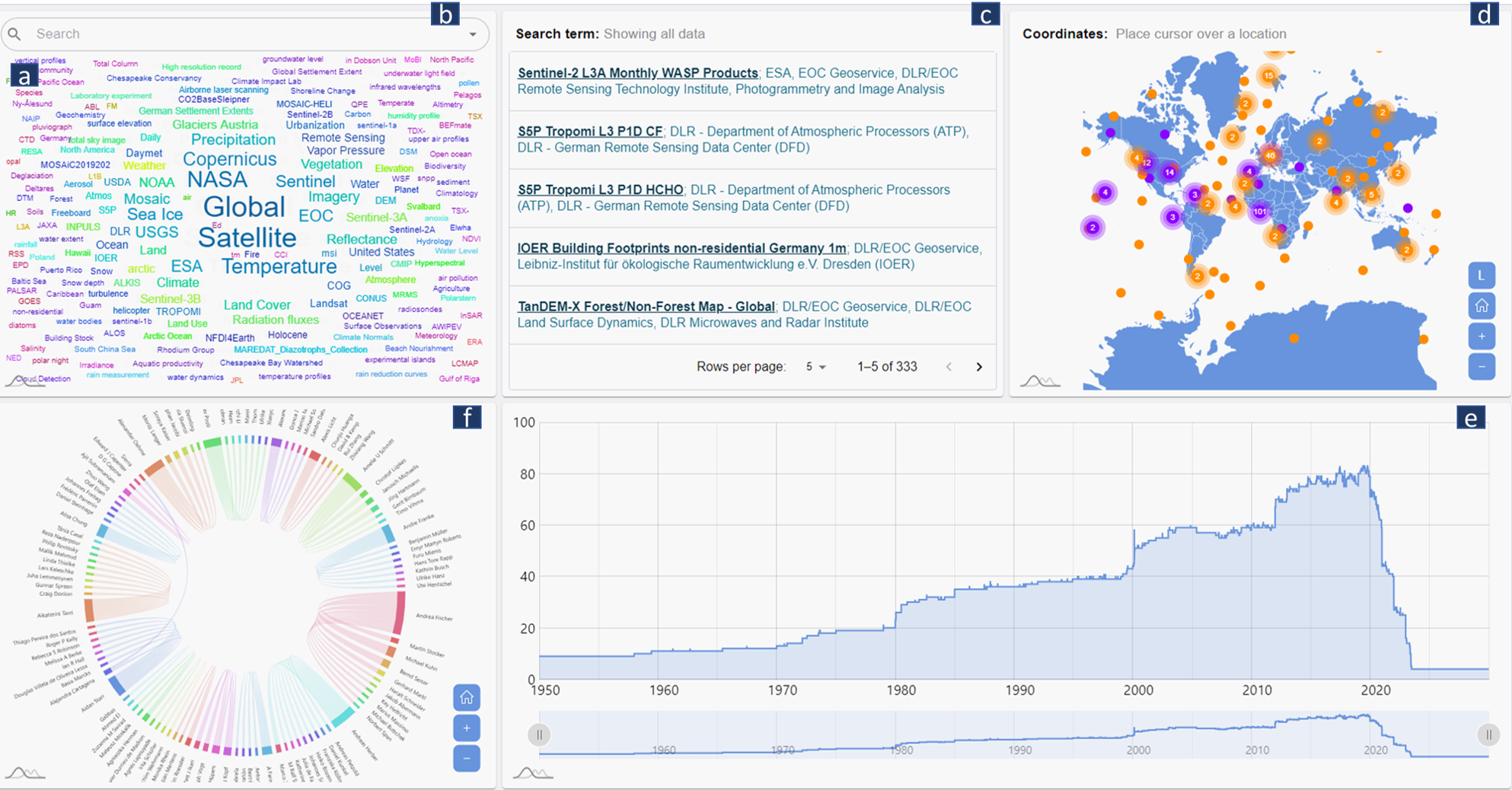}
  \caption{This is a frontend of our Knowledge Graph based Visual Search Application also named as VESA. This application helps to intuitively search for the datasets from currently two repositories i.e., Pangaea and DLR Earth Observation (DLR EO) data Center. Various visualizations helps in multidimensional data search. These components are: a) Wordcloud for contextual search, b) Autocomplete search bar also for contextual search, c) List showing search results and links to the data sources, d) Map for spatial search, e) Line Chart for temporal search and f) Chord Diagram to show relationship among common authors.}
 \label{fig:teaser}
\end{figure}

Above, we have highlighted four challenges in the traditional and current data search systems:
\begin{enumerate}
    \item Inability to search from multiple data repositories
    \item Heavily dependent on keyword and text based search
    \item Missing overview of the data collection
    \item Limited representation of implicit or explicit relationship among search entities.
\end{enumerate}
These challenges emphasize the need for improved data discovery tools, with user-friendly platforms that provide new concepts of discovery. In our efforts to address these issues, we developed a Knowledge Graph Based Visual Search Framework (\autoref{fig:framework}) and a software application VESA (\autoref{fig:teaser}) based on that. The backend of this application is a knowledge graph which harvest metadata from different repositories and connect their results based on the common and related attributes. At the frontend is a dashboard leveraging different visualizations to present search results from different perspectives.

For the Earth System Science (ESS) domain, we have developed a software prototype VESA (\autoref{fig:teaser}) based on this framework. It connects metadata from two repositories via a knowledge graph: PANGEA (\cite{felden2023pangaea}) and DLR Earth Observation (api available at \url{https://geoservice.dlr.de/eoc/ogc/stac/v1/collections/}. It allows the exploration of the collection and the search results on a visualization dashboard via four perspectives: Word Cloud for contextual, Map for spatial, Line Charts for temporal and Chord Diagram for inter-relationships (common authorship).

\subsection{Contribution}
In this paper, we introduce:
\begin{enumerate}
\item Our concept of a knowledge graph based visual search framework for data search environments. It \textit{(a)} facilitates data search from distributed data sources, \textit{(b)} introduces innovative way to present search results from different search perspectives, \textit{(c)} highlights implicit and explicit relationships within the metadata entities, \textit{(d)} provides an overview of dataset collection for better insights. We believe that this solution addresses the four challenges mentioned at the introduction section effectively.

\item A functional software prototype and a open source repository to support further development and community collaboration. Wherein developers can configure the backend APIs to harvest data from their own data repositories

\item Guidelines and requirements for designing effective search environments tailored for scientific datasets.
    
\end{enumerate}

We are certain that the above contributions advance the current state of the art in data search for scientific and research data management. The software tool is available at \url{https://vesa.webapps.nfdi4earth.de/} and the open source repository at \url{https://github.com/DLR-SC/VESA}.

The remainder of this paper is structured as follows. First in \autoref{section: related_work}, we reviewed the related work from different perspectives and examined the challenges identified in the empirical studies. Then, we introduce our proposed framework, detailing its key functionalities in \autoref{section: framework}. This is followed by two use cases that demonstrate its applications in two different scenarios in \autoref{section: use_case}. Then in \autoref{section: user_evaluation}, we present the results of our evaluation. In \autoref{section: guidelines}, we present the guidelines for the construction of an effective data search application. Finally, we discuss the challenges faced during development and propose future direction for VESA in \autoref{section: discussion} and provide conclusion in \autoref{section: conclusion}.
\section{Related Work}
\label{section: related_work}

Our framework builds upon existing research in data search services, user behavior, visualization systems, and knowledge graphs. Previous studies highlight key challenges in data discovery, including distributed data access, user sensemaking, and intuitive result exploration. In the following, we further examine these challenges and the available solutions.

\subsection{Challenges in Data Search Services}
Scientific data discovery has received increasing attention from both academic and infrastructure communities. This is reflected in the expansion of the coordinated efforts of organizations like the Research Data Alliance (RDA) (\url{https://www.rd-alliance.org/}), Go FAIR (\url{https://www.go-fair.org/}), OPEN AIR (\url{https://www.openaire.eu/}) initiative. All of which promote data sharing, interoperability, and the development of discovery-focused infrastructures aligned with FAIR principles. Despite that, most current systems continue to apply search paradigms originally designed for document retrieval, without accounting for the fundamental differences between documents and datasets. Several studies (\cite{kern2015there, kramer2021data}) have pointed out that dataset discovery involves unique challenges—such as interpreting structured metadata, handling incomplete or ambiguous coverage descriptions, and supporting exploration across spatial, temporal, and relational dimensions. Unlike documents, which can often be interpreted through title and full-text analysis, datasets require contextual understanding of their attributes, provenance, and use cases. Yet, the dominant approach in existing systems remains keyword-based querying, resulting in limited relevance, insufficient overview, and a frustrating user experience when large numbers of results are returned (\cite{dork2008visgets}). Khalsa et al. (\cite{khalsa2018survey}), in their survey of 98 repositories, found that the majority of data portals still rely on basic keyword search interfaces, which present results in long, paginated lists. These lists provide little insight into how datasets are distributed or related, and users are often left to click through each entry to assess relevance. Additionally, studies such as Wu et al. (\cite{wu2019data}) and Borst et al. (\cite{borst2020patterns}) emphasize the growing demand for domain-agnostic solutions that can bridge disparate repositories and facilitate cross-domain data discovery. They argue that current systems often mirror the structure and logic of the repository they represent. Despite widespread promotion of FAIR data principles, the lack of interoperability and semantic integration contributes to the fragmentation of research data and hampers its reuse.

\subsection{User Behavior and Sensemaking in Data Discovery }
Understanding user behavior in data discovery is critical for designing effective data search systems. Unlike traditional information retrieval, dataset search  is more cognitively demanding process. In a study conducted by \cite{kramer2021data}, researchers are shown to engage in iterative, exploratory search processes, often refining queries multiple times and frequently using filters to locate relevant datasets. Koesten et al. (\cite{koesten2021talking}) further emphasize the sensemaking challenges users face, noting that dataset discovery involves evaluating the context, provenance, and relationships of data attributes and entities. This highlights the need for tools that support users not only in finding datasets but also in understanding them. The importance of navigation and sensemaking is echoed in earlier work by (\cite{fox1993users, perez2015visual}), who demonstrate how visual navigation aids comprehension when users must move through large information spaces. In support to that, studies in human-computer interaction and information visualization suggest that coordinated multiple views (CMV)—where interacting with one visual component updates others—can enhance exploratory tasks and promote deeper engagement (\cite{shneiderman2003eyes, heer2012interactive}). These techniques enable users to move fluidly between broad overviews and focused details, a process often referred to as "overview first, zoom and filter, then details-on-demand", as coined by \cite{shneiderman2003eyes}.

\subsection{Knowledge Graphs for Search Applications}
In exploratory search, a user moves between different information artifacts guided by semantic connections between them. Consequently, graphs are a natural representation of such a search space, and in particular knowledge graphs have emerged as a powerful tool for integrating and exploring distributed data. In the field of biomedicine the DataMed is a system that facilitates dataset search through a knowledge graph of biomedical concepts from the Unified Medical Language System (UMLS) ontology (\cite{ohno2017finding}). Dug from \cite{WaldropEtAl2022} is a semantic dataset search tool for biomedical data. It utilizes ontologies and knowledge graphs of biomedical relationships (e.g. disease–exposure links) to expand queries and find relevant datasets that a purely keyword-based search might miss. 

The OpenAIRE graph dataset (\cite{ManghiEtAl2022}) is a domain independent knowledge graph that connects various research outputs, organizations, persons and projects. It can be considered as a general purpose backend for scientific search engines. Similarly 
The European Union’s unified open data portal, \textit{data.europa.eu}, uses a knowledge-graph approach to aggregate and enable search over tens of thousands of datasets from EU member states and institutions. By normalizing metadata to a common ontology (DCAT), the portal effectively creates a distributed knowledge graph of datasets – linking datasets to associated organizations, themes, formats, and other attributes.

\subsection{Visual Analytics for Data Discovery}
Visualization systems have proven to be of significantly valuable in information retrieval. Particularly by enabling users to identify previously unknown patterns and insights (\cite{saraiya2005insight}) within complex data collections. Despite its potential, the adoption of visualization or visual analytics tools in research data discovery remains limited. Especially, when compared to their more widespread use for data analysis, result presentation and data exploration in scientific domain (\cite{kaur2018biodiversity}). In these fields, tools often support rich, user-driven exploration through timelines, maps, topic networks and hierarchies, and multidimensional relationships. However, in the realm of research data management and scientific data repositories, visual search and exploration tools are still in infancy. There are some notable efforts, such as GFBIO's VAT at (\cite{beilschmidt2017vat}) provides a powerful environment for search of similar datasets in biodiversity domain. However, it is constrained to their own data repositories and specific schemas, limiting the broader applicability and scalability. Similarly, domain-specific systems such as ebird \cite{auer2024eod}, Ocean Data View at \url{https://odv.awi.de/} or Vertnet at \url{https://vertnet.org/} provide detailed visual querying and mapping capabilities, but only within specific data domain. Several other platforms—such as the ARIADNE  portal at \url{https://portal.ariadne-infrastructure.eu/} for archaeological datasets and Germany’s national Geoportal at \url{https://www.geoportal.de/}—provide interactive map views, timelines, or thematic filters to aid discovery. These tools allow users to browse datasets through different visual perspectives. However, the initial search process remains largely text and keyword-driven, often requiring users to formulate precise queries before any visual context is available. As a result, users are still reliant on knowing what to search for, which limits the potential for true exploratory discovery. Moreover, the visual views provided in these systems are often disjointed—meaning the different components (e.g., map, list, metadata details) operate in parallel and lack strong interactivity or coordination. Actions taken in one view (e.g., selecting a geographic region) may not dynamically update the others, reducing the effectiveness of the interface for sensemaking or multidimensional analysis. Thus, there remains a notable absence of visually-driven search platforms that allow users to explore datasets through seamlessly integrated, multi-perspective views, connected by underlying semantic relationships.

\section{Knowledge Graph Based Visual Search Framework}
\label{section: framework}
\begin{figure*}
\includegraphics[width=\linewidth]{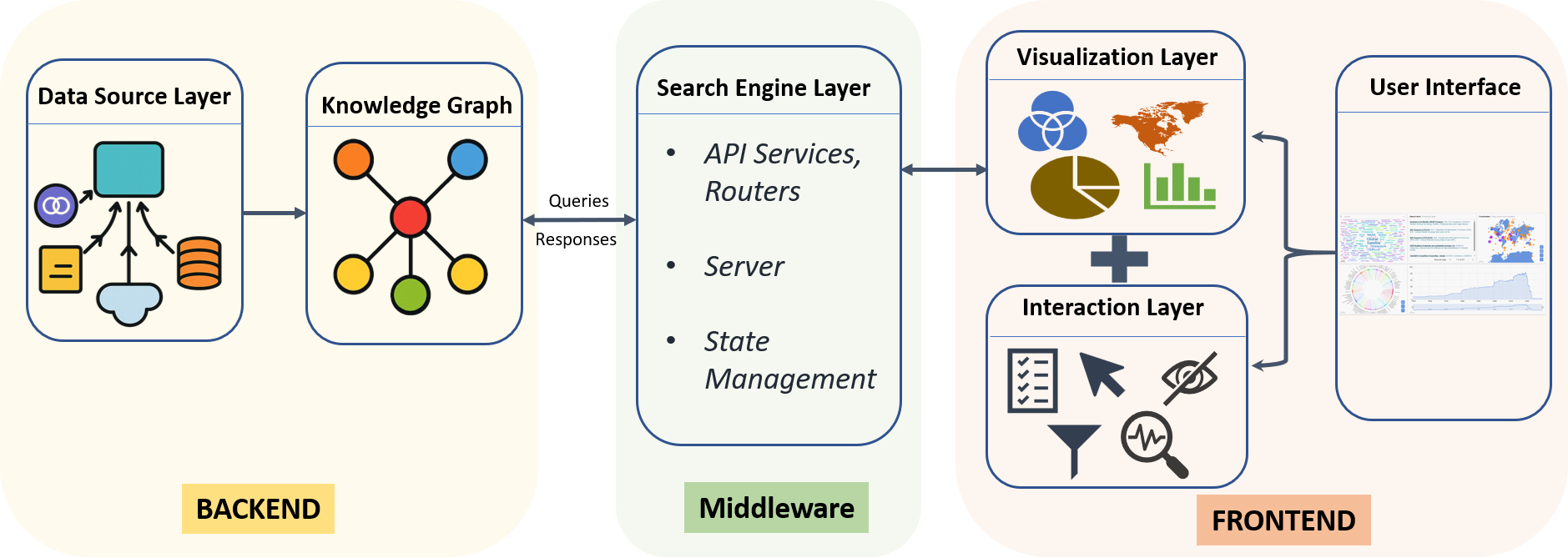}
  \caption{Our Knowledge Graph Based Visual Search Framework. It consists of three primary components: the Backend which serves as database, the Middleware which serves as search engine and the Frontend for showing the results to the users.}
  \label{fig:framework}
\end{figure*}

Our knowledge graph based visual search framework (\autoref{fig:framework}) presents a high-level architectural design of an intelligent data search system that supports access to datasets from multiple distributed sources and provides intuitive exploration of search results. The architecture of the system is organized into three primary components: the Backend, the Middleware and the Frontend. The Backend handles data ingestion, transformation, and semantic enrichment via a knowledge graph. The Frontend is responsible for user interaction, visualization, and query-driven exploration. The Middleware is responsible for providing all needed services for message passing and for stable data and web communication. In the following, we have further explained different layers of these components.

\begin{itemize}
    \item\textbf{Data Source Layer:} The Data Source Layer serves as the entry point of the architecture. It provides interfaces to connect and fetch metadata from multiple heterogeneous and distributed repositories—such as PANGAEA for geoscientific data and DLR EO for Earth observation datasets. These sources vary in format and structure. To ensure interoperability, the metadata is ingested, parsed, and normalized into a common schema. This harmonized data then becomes suitable for integration into a unified semantic structure.

\item\textbf{Knowledge Graph:} The normalized metadata is semantically enriched and structured into a Knowledge Graph, which is the central data model of this framework. Hosted in a graph database (e.g., ArangoDB), the graph captures entities like datasets, keywords, spatial and temporal extents, and authors as nodes. These nodes are linked via edges representing semantic relationships, such as "hasKeyword", "hasAuthor", or "coversLocation" (see VESA Knowledge Graph in \autoref{fig:kg}). The Knowledge Graph not only supports efficient querying and reasoning but also facilitates contextual exploration by interlinking concepts across domains.

\item\textbf{Search Engine Layer:} This layer act as a middleware and handles communication between the Frontend and the Backend components. It consists of API services and routers that receive structured queries from the interface and transform them into routes and valid calls to the the Knowledge Graph. The response is then filtered based on the queried data and sent to the Visualization Layer. 
The Middleware is also responsible for the overall State management of the application. By this mechanism, it keeps the application's behavior consistent during user interactions. This layer plays a pivotal role in ensuring low-latency, context-aware data retrieval and supports intelligent filtering based on user-defined criteria.  

\item\textbf{Visualization Layer:} Once data is fetched via the Search Engine Layer, the Visualization Layer transforms the response into formats compatible with visual components. Each visualization needs a data transformed into its specific format. For example for a Word Cloud it should have the information about the keywords, the related keywords, frequency of their occurance in the datasets and all dataset ids that have these keywords. The JSON pertaining to this looks similar to \autoref{fig:record}. In the same way, a Map needs information about the datasets and its spatial coordinates or regions and frequency of datasets in the specific region. Each visualization is then displayed by appropriate algorithms and design choices optimized for clarity and insight generation.

\item\textbf{Interaction Layer:} This layer brings the interactivity to the visualization and supports cross-filtering. Cross-filtering in visual analytics domain refers to an interactive data exploration technique where user selections or filters applied in one visual representation (e.g., a map or bar chart) are automatically propagated and reflected across multiple coordinated views. This enables users to explore multi-dimensional relationships and patterns (\cite{north2000snap, roberts2007State}). This layer provides all necessary tools to effectively explore and query the data directly through the interface: listings, clicks, filters, zooming, panning etc.

\item\textbf{User Interface:} The User Interface (UI) is the interactive space where users engage with visual components to explore data. Through interactive components, users can perform actions and immediately see the system’s response reflected across multiple coordinated views and visualizations. This direct feedback loop supports intuitive exploration and allows users to iteratively refine their queries based on visual evidence. Furthermore, the UI layer also serves as a gateway to external resources. Through the provided hyperlinks, users can navigate directly to the original data source websites or access the DOI links of datasets and related publications.
Additionally, the UI layer can be configure to provide other options for enhanced data discovery. It can be integrated with external analytical plugins to enhance data understanding. For example, users can link to plugins for quality assurance, raw data exploration, or generating summary statistics of the search results.
\end{itemize}
 
\subsection{Knowledge Graph Construction}

The main idea of the framework is to enable integrated visual search over items in heterogeneous data repositories by semantic interlinking of artifacts. To this end, metadata of data items as well as scientific publications are utilized to establish those connections. The resulting graph is hosted on an ArangoDB (\url{https://arangodb.com/}) graph database instance for use by the application frontend. The decision for this lightweight solution instead of relying on common knowledge graph technologies, such as RDF and SPARQL, was made because the graph only serves application-specific purposes and is not meant to be exploited to the web as is. This leaves flexibility to changes in the ontological structure and data types. However, elements in the graph which constitutes dataset items are associated with an persistent identifier (PID), if present, so that interaction with external semantic knowledge bases is still possible.  

For graph construction the Corpus Analytics Graph Builder (CAG) framework (\cite{el-baff-etal-2023-corpus}) is used. It automatizes mapping heterogeneous document corpora (in this case dataset collections) into a hierarchical graphical structure. On the top level there are so called "corpus nodes" representing a particular data corpus. Corpus nodes can be connected to other nodes representing corpus items, which may be further broken down into smaller pieces. On the bottom level there are typical corpus independent abstractions of information artifacts such as "text nodes" that represent textual content of a corpus item. 

\begin{figure*}
  \centering
  \includegraphics[width=\linewidth]{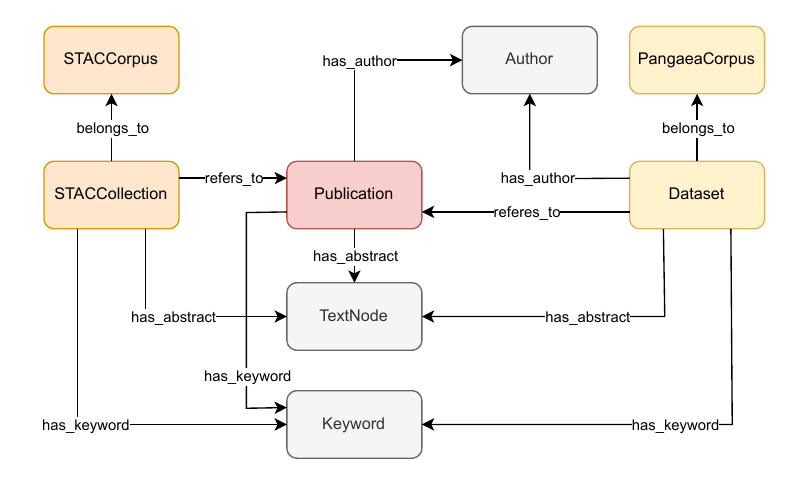}
  \caption{VESA Graph Structure. It shows how two different repositories are connected together.}
  \label{fig:kg}
\end{figure*}

The graph structure is depicted in Figure \ref{fig:kg}. As described before, VESA connects the dataset repositories PANGAEA and the DLR Geoportal. Pangaea datasets are directly mapped to nodes in the graph (yellow nodes in Figure \ref{fig:kg}). Dataset nodes store all available dataset metadata such as time information, geocoordinates, abstract, etc. and are connected to nodes representing the authors of the dataset. For authors typically only the name and organization are available. Furthermore, some Pangaea items give explicit references to related publications, which are also stored in the graph database. 

In the DLR geoportal, earth observation data is stored in the form of Spatio Temporal Asset Catalogs (STAC) (\url{https://stacspec.org/}), a standardized format to store and access geospatial information. Each STAC collection is mapped to a node in the graph storing the specific catalog metadata. 

Both data portals differ significantly in their nature and organization. Semantic connections between data items from both sources are established based on common keywords. While datasets in Pangaea are directly associated with a curated list of keywords as part of their metadata, this is not always the case for STAC catalogs. In order to assign keywords to STAC catalogs that provide not enough metadata, publications are used as mediating artifacts. In earth observation related publications it is common practices to mention the earth observation mission, which data was used in the paper. In this way STAC catalogs can be indirectly connected to the other keywords extracted from the metadata of the paper and thus be related to Pangaea datasets. For example, if a paper mentions the earth observation mission TerraSarX and has an additional keyword "flood events", the corresponding STAC collections can be linked to other resources that also share the "flood events" keyword.

In this way over time more and more semantic relationships between datasets can be added to the database, which constitutes a rich source of information and allows for the discovery information that are not available from one dataset alone.

\subsection{Visualization Enabled Search Application (VESA)} \label{section:vesa}

The design of VESA interface is inspired by the work of Marian Dork in \cite{dork2008visgets} and previous work related to multiple linked views (MLV), facet search, visual querying, and exploration on the web.

Unlike traditional text based search interfaces, VESA leverages different visualizations to provide efficient, robust and multidimensional search results. The exploration logic in VESA is based on question queries that typically starts with "W" and ends with "?" in text queries. Different studies (\cite{Aula2010HowDS, Tamine2010EvaluationOC}) have analyzed the effectiveness of "W" question queries in search process, demonstrating that they enhance search accuracy and improve retrieval efficiency. This also holds true for observation datasets, which provide information about the observed entity and the context of the observation (\cite{madin2008advancing}) (e.g. when and where the observation was taken and who recorded it). Such contextual information is typically embedded within the dataset’s metadata, as shown in this reference (\url{https://doi.pangaea.de/10.1594/PANGAEA.958479}). 

For using VESA for ESS, we used six metadata attributes from the knowledge graph. They are: \textit{Title, Abstract, Keywords, Authors, Spatial Coverage, and Temporal Coverage}. 
\textbf{Word Cloud} (\autoref{fig:teaser}a) is linked to the keywords extracted from Title, Abstract and Keywords attributes of the metadata. The size of the word shows the frequency of that term referred in different datasets. Thus higher the frequency, more prominent that term is in the collection. Selecting one keyword further shows the related keywords in the Word Cloud and consequently filters the other related attributed and datasets in the connected visualizations. In order to show only relevant and important keywords, their weights were derived via TF/IDF metric, where the number of documents are the number of metadata files in which the keyword is available. Based on the weights of TF/IDF, keywords with high scores were then filtered and displayed in the visualization. Thus not all keywords were shown from all 333 datasets. However, a provision is made to search for all keywords in our data store, via running a search from the \textbf{Autocomplete Search Bar} (\autoref{fig:teaser}b) component. \textbf{Map} (\autoref{fig:teaser}d) enables filtering based on the spatial coverage of the datasets. Currently two data repositories are shown, depicted with dots of two different colors. Hovering over the dot also shows its coordinates on the top bar of the map canvas. The dots or points on the Map get automatically clustered when a user zooms out of the Map or when the Map is at its default position. Then on the clustered ball user can see the number of datasets related to the clustered region. At the left corners different icons in blue colors are placed. 'L' is to enable and disable Legend, Home icon brings the map back to its default position, '+' is for zooming in and '-' is for zooming out of the map. \textbf{Line Chart} (\autoref{fig:teaser}e) shows the temporal coverage of the datasets. The default X axis shows the temporal range which is defined by the maximum and minimum date range of all the datasets available in our data store. The Y axis shows the number of datasets. The Line Chart below helps in zooming in and filtering through the years and then the result is shown on the Line Chart above. From here, one can filter dataset based on years, months and dates. Hovering over the line on this chart shows the number of the datasets belong to the specific date. The \textbf{Chord Diagram} (\autoref{fig:teaser}f) shows the frequency of the co-authorship of the filtered datasets. The thicker a chord is the more frequently two authors have contributed in the same dataset. A \textbf{List View} (\autoref{fig:teaser}c) in the middle presents the filtered results. It shows the title and the authors of the filtered datasets. Clicking on the title will take user to their actual DOI or source page. This view also shows the count of the retrieved datasets. 

\subsection{Search Engine Layer of VESA }
\begin{figure}

  \includegraphics{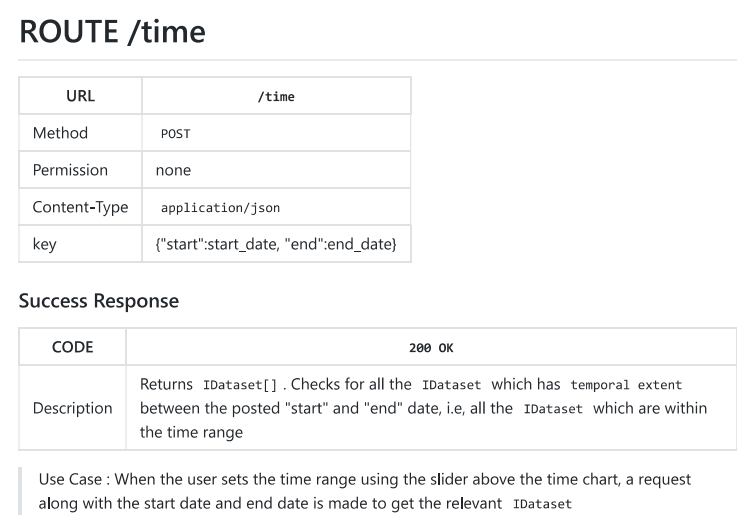}
  \caption{An example of a route defined. /time route gets metadata information of all the datasets which are present between a given time range.}
  \label{fig:routes}
\end{figure}

VESA's search engine layer (refer \autoref{fig:framework}) runs on a Node.js (\url{https://nodejs.org/en}) server and provides different APIs  or routes for fetching data from the knowledge graph. The data from these routes are then processed and sent to the Frontend. For bidirectional message communication between Frontend, Middleware and Knowledge Graph different routes are developed which further runs ArangoJS queries. When a user clicks on some visualization, the route connected to it gets activated and sends a request to the server to get the required response. An example of one of the route is shown in \autoref{fig:routes} for Line Charts.
There are five such routes for VESA:
\begin{enumerate}
    \item \textit{/main/all:} This route get all the metadata information of the dataset and the publications: list of
authors, spatial co-ordinates, temporal information, title, dataset id, doi etc. A sample response is provided in the Listing 1. This information is then shown via respective visualizations of the VESA interface. For example \textit{location\_data} to Map view, \textit{temporal coverage} to Line Charts etc.
\item \textit{/keyword:} This routes gets the keyword information like the keyword name, its TF-IDF score, and the dataset ids in which the keyword is actually present. This information is then shown in the Word Cloud (\autoref{fig:teaser}a).
\item \textit{/time:} This is used to get all the Datasets which are present between a given time range. This information is then shown in the Line Charts (\autoref{fig:teaser}e).
\item \textit{/abstract:} This is used to get the abstract of a Dataset. This information is then shown in the List view (\autoref{fig:teaser}c), when user clicks on any row.
\item \textit{/map:} This is used to get spatial co-ordinates of each dataset.
\end{enumerate}

\begin{figure}[h]
\includegraphics[width=\columnwidth]{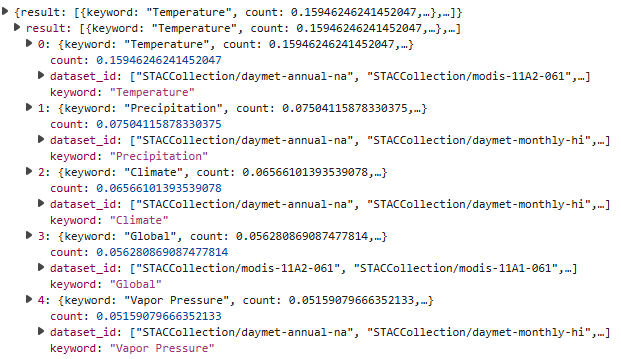}
  \caption{Resultant record when a 'Temperature' keyword is clicked on the Wordcloud.}
  \label{fig:record}
\end{figure}

During the first load or everytime the application reloads \textit{/main/all} and \textit{/keyword} gets activated. \textit{/keyword} also gets triggered when a user select any keyword from the Word Cloud. This route further activate functions to get all the related keywords for the selected keyword. In \autoref{fig:record}, we show a resultant record when a 'Temperature' keyword is clicked. All other related keywords are fetched from the \textit{dataset\_id} which has this keyword.

\begin{lstlisting}[basicstyle=\small\ttfamily,  breaklines=true]
"result": [  
{
    "id": "Dataset/495977132",
    "location_data": {
        "west_bound_longitude": -58.0365,
        "east_bound_longitude": -45.688,
        "north_bound_latitude": 61.4639,
        "south_bound_latitude": 50.208,
        "mean_latitude": 56.62752222222233,
        "mean_longitude": -50.69916666666666
    },
        "doi": "https://doi.org/10.1594/PANGAEA.958142",
        "dataset_publication_date": "2023-11-13T06:33:47+00:00",
        "temporal_coverage": {
        "start_date": "1999-07-31T23:00:00Z",
        "end_date": "1999-08-01T23:00:00.000Z"
    },
        "authors": [
            "Franziska Tell"
    ],
        "dataset_title": "Individual shell sizes 
        and shell weights of planktonic foraminifera 
        from five samples from the Labrador Sea 
        cores HU2008-029-004TWC, HU91-045-93BX and MD99-2227",
        "organization": "PANGAEA"
        }
Listing 1: JSON output for /main/all response.
\end{lstlisting}
\label{lst:main}

\section{Use Cases}
\label{section: use_case}
In the following, we illustrate the relevance and applicability of the proposed system through two use cases from two different users of a data search system.

\subsection{User Searching for a Dataset} 
\begin{figure*}
  \centering
  \includegraphics[width=\linewidth]{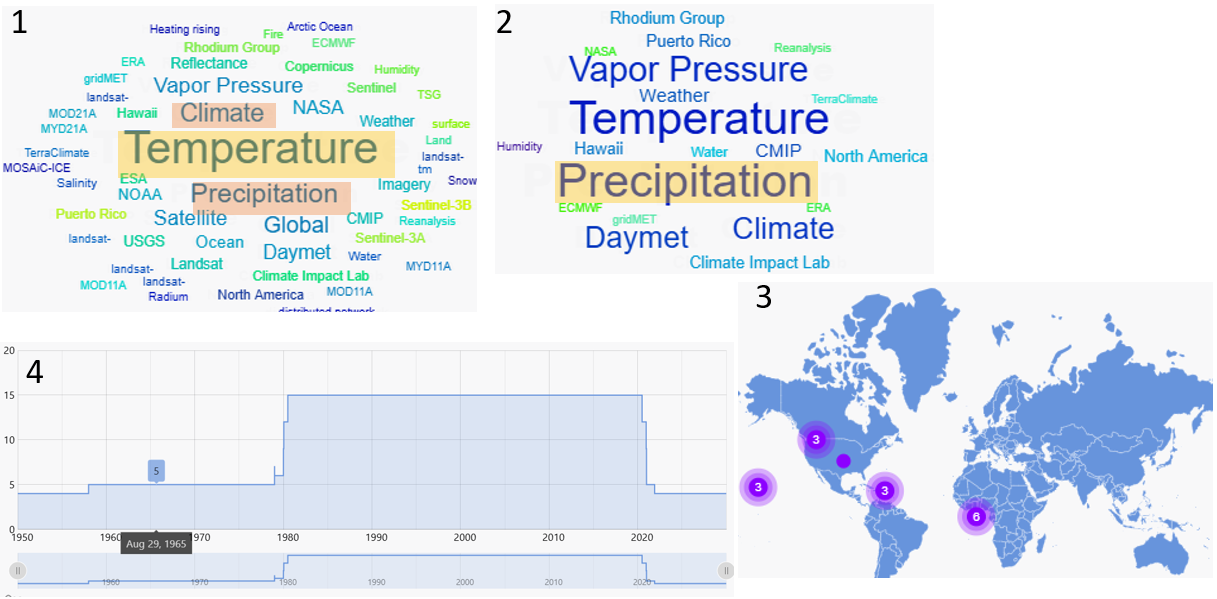}
  \caption{Screenshot from running the use case 1 on VESA.} 
  \label{fig:usecase}
\end{figure*}

Consider a scientist named Dr. Smith researching climate change impacts. He might
start by clicking on a keyword ’temperature’ into the Word
Cloud visualization. This action would show other related keywords
such as ’climate’ and ’precipitation’, and filter the datasets accordingly across the other visualizations (see \autoref{fig:usecase}-1). Selecting a keyword ’precipitation’ would further refine the results to show datasets related to
precipitation measurements on the Map, the relevant time periods
in the Line Charts, and the authors involved in these studies in the
Chord Diagram (see \autoref{fig:usecase}-2,3,4) respectively. By this iterative process, a user can efficiently refine their search and can quickly identify the most relevant datasets without the need for multiple separate queries. This streamlined approach saved Dr. Smith time, eliminating days of manual search and enable him to compile a comprehensive and related dataset for his research.

\subsection{Decision Maker (Data Manager, Principal Investigator or Funding Institutions} A Funding Institution (FI) seeks to assess the extent of past research conducted on ocean studies. To achieve this, they use VESA. By clicking on 'Ocean' in the Word cloud, they first gain insights into the various research themes related to ocean studies. The related keywords indicate connections to Ocean Temperature, Geochemistry, Climate, Nutrients, and Global observations (which are not linked to specific geographical locations). From the List View they know that in total 10 datasets are produced with their funding. Out of which two are stored in PANGAEA repository and eight at DLR Geoportal.  Moreover, they observe a notable increase in dataset production starting in 1970, peaking between 1988 and 2020, after which no further observations were recorded. Furthermore, their exploration identifies two key scientists who have played a major role in generating these datasets.  This information helps decision-makers evaluate past research efforts, identify areas of stagnation, and make informed decisions on future research directions and funding priorities.

\section{User Evaluation}
\label{section: user_evaluation}

To assess the usability and effectiveness of our system, we conducted a evaluation using an online questionnaire (see supplementary material). The survey was designed using Google Forms and aimed to capture both quantitative and qualitative feedback from domain experts and potential end-users. Those in our case were researchers or scientists in the field of ESS or other physical science domains. The responses were taken anonymously and participants were asked to only identify their scientific domain, age group, and frequency of interaction with data visualizations. 

To evaluate participant performance in interpreting the interface, the questionnaire included a short experimental task. Participants were asked to read specific information from four different visual components of the interface. Their responses were then compared against the correct reference values. The number of errors made by each participant was recorded, with fewer errors indicating higher performance. These correctness-based tasks are a commonly accepted method for objectively assessing user performance in visualization and interface studies, as they provide measurable evidence of how effectively users can interpret and interact with visual elements (\cite{purchase2012experimental}).

Further open ended questions were focused on whether the visualization components facilitated deeper insight, triggered further exploration, or led to the formulation of new queries. Participants were also asked to compare their experience with VESA to other data search tools they typically use. The questionnaire concluded with five-point Likert scale evaluating the overall perception of the system across dimensions such as intuitiveness, clarity, design aesthetics, ease of use with limited training, and good in capturing interest.  

\subsection{Results}
\begin{figure}[t]
  \includegraphics[width=\columnwidth]{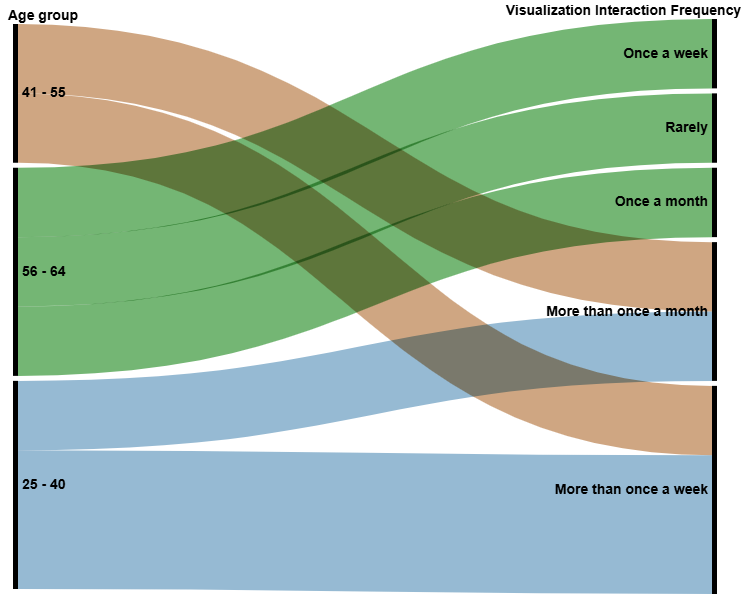}
  \caption{A relationship between age group and data visualization interaction frequency.}
  \label{fig:frequency}
\end{figure}

In total eight participants took part in the survey, with the age groups vary primarily from 25-64 years. Data visualization interaction frequency ranges from "Rarely" to "More than once a week," suggesting different familiarity levels with data visualization tools. As shown in \autoref{fig:frequency}, age groups under 55 years users are more frequently using visualizations for their work (more than once a week and month).


\subsection{Participants performance with the experimental task}
Due to the inherit nature of the survey, no prior training or walkthrough was possible. Therefore, it was essential to assess the correctness of the responses in order to determine whether participants were able to independently understand and interpret the visualizations, regardless of their prior experience with such charts. The evaluation focused on four visual components: the word cloud, list view, line chart, and map view. Results showed high levels of accuracy across all views: 100\% correct responses for the word cloud, list view, and line chart, and 70\% correctness for the map view. In total, only one error was observed across 32 individual responses, indicating a strong overall understanding. Thus, the participant performance in this task was approximately 97\%, demonstrating that the visual components of the interface were largely correctly interpreted without requiring prior training.

\subsection{Comparison with other tools}
Our participants use varied data search applications where majority were using PANGAEA and Google Dataset search and others were BASE, DATACITE SEARCH, WDCC, ECMWF, R and interfaces from their own repositories. When asked to rank our search application in comparison to their usual one, then 50\% consider them better, 38\% consider it same and 12\% consider it worst. Further for the questions on \textit{"If the overview of the datasets from the two repositories is clear to understand and intrigued any new insight or further search queries?"} The reaction was mixed. Some appreciated the visual exploration and noted interesting trends (e.g., dataset growth over time) and knowing the relationship between authors and publications. Others found the interface initially overwhelming and requested more onboarding or guided navigation. 

\subsection{Usability evaluation}
\begin{figure}

  \includegraphics[width=\columnwidth]{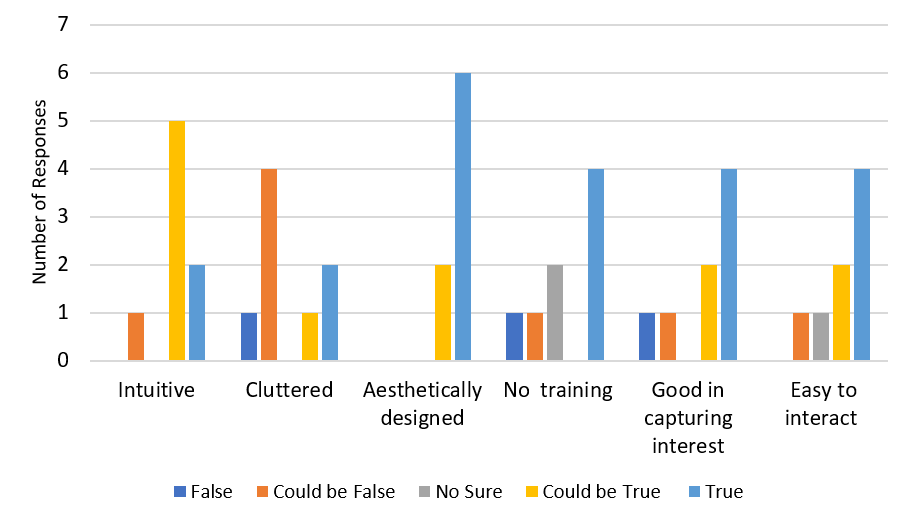}
  \caption{Results on usability of the VESA interface. The clustered bar chart shows user perception across six key dimensions:  intuitiveness, interface being cluttered, aesthetically designed, no training needed, good in capturing interest and easy to interact with. }
  \label{fig:usability}
\end{figure}

Further we summarized the results of the Likert-scale questions related to the expressiveness and usability of the user interface (see \autoref{fig:usability}). Majority participants consider this tool as \textit{'aesthetically designed'}, which needs \textit{no to limited training}, is \textit{good in capturing the interest} and \textit{is easy to interact with}. On the topic of application being cluttered or not, participants had mixed reaction. \autoref{fig:usability} shows that around 50\% answered \textit{it could be false}, however comments pointed word placements in word cloud to be cluttered and could be fixed. In word cloud, algorithm place the terms randomly based on the frequency of its occurrence. On the other hand, some considered this visualization intriguing and somewhat intuitive. When tried to understand the reason, we realized that many participants had major issues with no description, tutorial or help available for the tool. As one said \textit{"as first time user, I dont get sense of the lower two frames. I would need more labels and explanations."} Moreover, as this tool was based on limited data sources and datasets, many were not able to find their datasets leading to further more doubts. 


\subsection{Qualitative analysis of the user responses} Overall it was a positive response. As one mentioned \textit{"it is a modern approach and enable a play factor, which keeps interest."}. Other said \textit{"It certainly looks intriguing. The keywords are nicely displayed. The interaction of the authors shown."} They liked to drill down the search through different aspects and can visually see the result. Also liked the overview of available datasets at a single glance. However through out their comments, it is found out that some level of training is required either in the form of manual or a tutorial. Further we were able to extract some areas of improvement, which are: adding more datasets and data sources, options to rearrange different frames, a color pallette. Some issues were related to the missing data and metadata quality as one said \textit{ "I get a landsat result with a location shown in the map - landsat data  should be over a region."}

\subsection{Summarising Results}
In \textbf{summary}, the survey results indicate a generally positive reception of our system among participants from diverse scientific domains. Most users found the system to be aesthetically designed, easy to interact with, and capable of capturing interest. Despite no prior training, participants performed remarkably well in understanding the visualizations, with a 97\% overall correctness rate across four visual components—word cloud, list view, line chart, and map. Qualitative feedback highlighted appreciation for the visual exploration features and interface design, while also pointing out areas for improvement such as clearer labeling, help documentation, and reduced complexity for first-time users. In comparison to their usual search tools, the majority rated the application as either equal to or better, suggesting strong potential for adoption, particularly with minor usability enhancements. These findings support the system's effectiveness as a visual search interface and provide a strong foundation for future iterations. 
However, it is important to note that the findings are based on a limited sample size and number of datasets (333 datasets), which may constrain the generalizability and statistical significance of the results. While the insights provide encouraging early evidence of the system’s usability and effectiveness, a large-scale user study would be necessary to draw more definitive conclusions.

\section{Guidelines for the development of an effective data search system}
\label{section: guidelines}
In the following, we have summarized and listed some of the requirements for an effective data search tool for the research data management environment. These guidelines are derived from the analysis of the literature and our experience with VESA. These requirements could be considered as guidelines before one starts in building search systems for data management.

\begin{enumerate}
    \item \textbf{Support for Lookup, Exploratory and Overview Tasks:} Effective data search tools must cater to both specific lookup tasks and open-ended exploratory tasks, as also highlighted by \cite{kramer2021data}. Users should be able to locate precise datasets efficiently (lookup tasks) while also having the capability to explore broader patterns, trends, and relationships within the resultant filtered view (exploratory tasks). Apart from that, studies have indirectly emphasized the need of showing comprehensive overview of the collection. This further helps orientation and context-building from the filtered search space (\cite{koesten2021talking}). As shown in \autoref{section: use_case} second use case,  such overviews are also critical for data managers, institutions, and funding agencies for repository evaluation and strategic planning.

\item \textbf{Metadata Integration and Interoperability:} Effective search environments should provide a provision to integrate metadata from diverse and distributed repositories seamlessly (\cite{borst2020patterns, mertz2022neocube}). This means handling different types of metadata schema and making them work together. Using ontologies to map metadata fields ensures that similar concepts from different sources can be searched together. By doing so, search environments can facilitate metadata interoperability, enabling seamless cross-repository searches and enhance connectivity between datasets across different domains.

\item  \textbf{Multidimensional Search and Visualization:}
Search application should enable search via different concepts (\cite{khalsa2018survey}) and should present search results across multiple dimensions (\cite{blazevic2021visual}). Although the selection of dimensions is very domain specific, there are always some common dimensions in each domain that can be used to create a base system. For example, studies have tried to understand some basic dimensions for observation datasets (\cite{madin2008advancing, otegui2012biddsat}),  which for our work we identified as spatial, temporal, hierarchical, and contextual. This list can further be altered and enhanced based on the specific data domain. Semantic technologies can further enhance this list by automatically categorizing the search dimensions from the metadata and datasets. Interactive visualizations techniques can then assist in showing these dimensions via various chart types. 

\item \textbf{Domain-Agnostic Functionalities:}
Search tools must be scalable enough to integrate many repositories and should support interdisciplinary research without being constrained by the specificity of individual domains (\cite{wu2019data}). For which domain agnostic tools are needed that can seamlessly connect multiple repositories as per user demand and can effectively display the result at the frontend. It needs a robust and highly configurable system, which user can edit without much complexity. At backend, via use of different knowledge graph endpoint, different repositories can be connected. At the frontend, provision should be given to include different visual metaphors and chart types.

\item \textbf{User-Centric Design and Intuitive Interfaces:} 
Even the most powerful tool is ineffective if people can’t figure out how to use it. Data Search tools must prioritize usability and intuitive interfaces (\cite{nowell1997exploring, jeong2011investigation}). Features like autocomplete search bars, linked visualizations, and cross-filtering reduce cognitive load. Furthermore labeling, captioning and contextual narratives throughout the interface can play a crucial role in growing user understanding. Moreover, built-in guidance, such as tooltips, onboarding wizards, or walk-through tutorials, can lower the learning curve. These features are especially helpful in complex search environments where users may need to understand multiple filters or interactive components. 

\item \textbf{Support for Sensemaking:} 
Sensemaking is about helping users understand and interpret what they find. Tools must assist users in interpreting and deriving meaning from search results (\cite{angelini2017visual}). This requires integrating features that reveal relationships, summarize search results for easier comprehension, highlight key attributes or even automated suggestions.

\item \textbf{FAIR Principle Alignment:} Search tools must adhere to the FAIR data principles  — Findable, Accessible, Interoperable, and Reusable. For modern search systems: particular emphases should be given to the aspect of findability, interoperability, and reusability. This includes enabling users to locate datasets easily, ensuring metadata compatibility across repositories, and providing clear links to dataset sources for reuse.

\item \textbf{Evaluation and Feedback Mechanisms:}  
Tools should include mechanisms for evaluating their effectiveness and incorporating user feedback. In (\cite{seebacher2017visual}), author emphasize that iterative evaluation and adaptation improve the alignment of tools with user needs. This is possible via adopting a participatory design approach (\cite{janicke2020participatory}) with iterative development and regular feedback sessions.
\end{enumerate}

\begin{table}[h!]
\centering
\begin{tabular}  { |p{5cm}|l|p{9cm}| } 
     \hline
   Requirement  & Score & Explanation 
    \\  \hline

Support for Lookup, Exploratory and Overview Tasks & 1 &  VESA provides multidimensional exploration and overview of the datasets in store. It assists in lookup and exploration of the filtered datasets. \\  \hline 
Metadata Integration and Interoperability & 2 & Currently it only support two data repositories and connects on specific metadata fields. \\  \hline 
Multidimensional Search and Visualization  & 1 & It provides visual search in four dimensions. \\  \hline 
 Configurable and Domain-Agnostic Functionality & 3 & Currently, neither VESA's backend or frontend is domain agnostic.\\  \hline 
User-Centric Design and Intuitive Interfaces  & 2 & VESA has an intuitive design, however, more user studies needs to be done to make it more user-centric. \\  \hline 
Support for Sensemaking & 2 & Currently, it only support Lookup, Exploration and Overview tasks and does not summarize the search datasets. \\  \hline 
FAIR Principle Alignment & 2 & This tool does not provide metadata compatibility as it is tailored for only two repositories. However, it helps in finding the datasets and provides the source data link. \\  \hline 
Evaluation and Feedback Mechanisms Tools  & 3 & No online and live evaluation and feedback mechanism are built in it. Though, with every major release the tool will be evaluated manually from its users. \\  \hline 
    \end{tabular}\\ 
    \caption{Evaluating VESA based on the derived guidelines. We scored it 1, 2 and 3 based on if it fulfills all, some and none of the tasks in each requirement.}
\label{tab:development}
\end{table}

Further, we present \autoref{tab:development} that shows how VESA fulfill the derived requirements presented above. The table shows that except two points - 'Domain agnostic' and 'Evaluation of feedback', it fulfills partially to all the tasks within the derived requirements. As currently it is only deployed for two repositories of ESS, it could not fulfill the demands of other metadata standards (requirement 2). To make it more user centric and configurable, we need more evaluations and more use cases. Furthermore, for summarization of the datasets, a separate tool is needed which then would be integrated with VESA.

\section{Discussion}
\label{section: discussion}
For this work, we combined the knowledge graph technology with visualization dashboards, as a solution to  solve some of the problems data search systems are facing in the modern days : distributed search, overview of the collection, interactive and better search efficiency and showing implicit and explicit relationships among datasets. Thus providing a first use case of visualization enabled search environments for research data management. 
One of the major contributions of this framework, lies in its ability to harmonize data from different repositories through the use of a knowledge graph. This architecture not only supports FAIR principles but also provides a foundation for semantic search and relational exploration. The dynamic interaction between visual components demonstrates the tool's ability to adapt to user queries iteratively, enabling an intuitive exploration. During the course of this work, we realized that only visualization-enabled search applications with coordinated views (such as VESA) are capable of providing a comprehensive overview of the entire resultant search space. This capability helps users derive deeper insights from the retrieved datasets. In contrast, text-based interfaces fail to support such tasks effectively.

Overall, the evaluation showed positive reception proving it as a modern and enhanced search system. However, the results must be contextualized within the limited number of datasets for the software and limited participants. One of the important requirements from the discussion with the community members were to make it more data agnostic. Some data managers were highly interested in directly deploying the frontend on their repositories to enhance their search functionalities. Some others were interested to add their repository to our existing knowledge graph. The framework demonstrates significant promise not only for Earth System Science but also for other disciplines that rely on distributed, heterogeneous datasets. Its application in research data management could serve as a model for integrating FAIR principles into practical, user-focused tools.

While the knowledge graph successfully establishes relationships between datasets, its reliance on metadata quality is a limitation. Inconsistent or incomplete metadata can lead to missed connections or irrelevant search results. Additionally, the feedback loop between user interactions and query refinement, requires a more robust backend to handle complex queries efficiently at scale. This opens up  future work for VESA related to the scalability of the knowledge graph and interaction at the frontend. Further work include provision to add more data sources to the knowledge graph, adding sense making functionalities, and making provisions to configure the frontend and visualizations based on the user feedback.



\section{Conclusion}
\label{section: conclusion}
This paper presents the Knowledge Graph based visualization enabled search application (VESA), a novel framework designed to address the challenges of modern data discovery. By leveraging knowledge graph technology and integrating multidimensional visualizations, VESA provides an intuitive and interactive platform for exploring complex datasets across multiple repositories. Its design not only aligns with the FAIR principles but also demonstrates the potential to redefine how researchers interact with and extract value from distributed, heterogeneous data sources. The software implementation, connecting two Earth System Science repositories, highlights VESA’s ability to facilitate relational, spatial, temporal, and contextual exploration. User evaluations revealed positive reception, emphasizing the framework's ease of use, low learning curve, and ability to uncover meaningful relationships within data. The study also identified areas for improvement, including the need for enhanced metadata quality, expanded dataset integration, and better support for user onboarding.

Future work will focus on expanding its applicability for various data domains, incorporating additional repositories, addressing scalability challenges and making the tool user configurable. By further refining its interactive capabilities and enhancing semantic integration, VESA brings to the community a transformative tool for interdisciplinary research and knowledge discovery.

\section*{Supplementary Files}
They are attached in a zip folder with this submission

\section*{Reproducibility}
Open source repository of this work is available at https://github.com/DLR-SC/VESA

\section*{Acknowledgment}
This work has been funded by the German Research Foundation (DFG) through the project NFDI4Earth (DFG project no. 460036893, https://www.nfdi4earth.de) within the German National Research Data Infrastructure (NFDI, https://www.nfdi.de). We would like to thank our student assistant Anirudh Panchangam Ranganath and Malikathazham Hudaif during the development of this software. We would also like to thanks all participants of our user evaluations and all those who provided their feedback for the enhancement of this tool.

\bibliographystyle{agsm} 

\bibliography{main}


\end{document}